# Integrated Metasurfaces on Silicon Photonics for Emission Shaping and Holographic Projection

*Ping-Yen Hsieh,[1] Shun-Lin Fang,[1] Yu-Siang Lin,[1] Wen-Hsien Huang,[2] Jia-Min Shieh,[1,2] Peichen Yu,[1] and You-Chia Chang[1,*]*

[1]Department of Photonics, College of Electrical and Engineering, National Yang Ming Chiao Tung University, Hsinchu 30010, Taiwan

[2]Chip Integration Technology Department, Taiwan Semiconductor Research Institute, Hsinchu 30078, Taiwan

ABSTRACT: The emerging applications of silicon photonics in free space, such as LiDARs and quantum photonics, urge versatile emission shaping beyond the capabilities of conventional grating couplers. A platform that offers arbitrary shaping of free-space emission while maintaining the CMOS compatibility and monolithic integration is in pressing need. Here we demonstrate a platform that integrates metasurfaces monolithically on silicon photonic integrated circuits. The metasurfaces consist of amorphous silicon nanopillars evanescently coupled to silicon waveguides. We demonstrate experimentally diffraction-limited beam focusing with a Strehl ratio of 0.82, where the focused spot can be switched between two positions. We also realize a meta-hologram experimentally that projects an image above the silicon photonic chip. This platform can add a highly versatile interface to the existing silicon photonic ecosystems for precise delivery of free-space emission.

Silicon photonics has recently shown promises in delivering free-space emission to detect or manipulate external objects with multifunctional photonic chips. Notable applications include light detection and ranging (LiDAR),[1–7] free-space optical communication,[8,9] display,[10,11] optogenetics,[12,13] and quantum computation.[14,15] Silicon photonic platform offers scalability, reconfigurability, compactness, parallelism, and mature fabrication, thanks to the complementary metal-oxide-semiconductor (CMOS) compatibility.[16] Silicon optical phased arrays, for example, have leveraged the scalability to integrate hundreds or thousands of phase shifters to form and steer free-space beams, offering the key functions for LiDAR and free-space optical communication.[1,3,4] In optogenetics and trapped-ion-based quantum computation, silicon photonics offers precise light delivery to the neuron or ion positions as well as the scalability to increase the number of addressable neurons and qubits, which is challenging for conventional bulk optics.[12–15] These emerging applications urge the generation of precise and sometimes complicated free-space spatial modes, including beam focusing,[14,17] collimation,[18,19] orbital angular momentum generation,[20] and holographic projection.[11] Beam focusing allows optical excitation or sensing exclusively of a small volume in space, which is crucial for addressing qubits in quantum photonics and neurons in optogenetics.[14,21] Generating a large-aperture collimated beam is the key to extending the ranging distance in LiDARs.[18,19] Beams carrying different orbital angular momentums can enable spatial-division-multiplexing for free-space optical communication.[20]

The emerging free-space applications of silicon photonics urge versatile emission shaping beyond the capabilities of conventional grating couplers and inverse tapers.[22,23] Although grating couplers can be apodized with a spatially varying period to perform beam focusing,[14] generalization to arbitrary shaping is challenging. To address the need for arbitrary free-space beam shaping, metasurfaces have been introduced to silicon photonics. Metasurfaces are planar arrangement of subwavelength-spaced optical scatters, the so-called meta-atoms.[24,25] Metasurfaces can impose local modulation on the phase, amplitude, polarization, and even dispersion of electromagnetic waves by assigning the geometrical parameters of each scatter.[24–28] While there has been a vast amount of research using metasurfaces as free-space components,[29–31] the integration of metasurfaces on photonic integrated circuits has yet been fully explored.[2,17,32–38] In the works by Chang *et al*. and Yulaev *et al*.,[2,17] silicon metasurfaces are placed on top of grating couplers to either collimate or focus the grating emission. However, because the metasurface and the photonic integrated circuit are fabricated separately, precise alignment and packaging are required. Monolithic integration of metasurfaces on silicon photonic integrated circuits has been proposed numerically[32,33,38] and demonstrated experimentally.[34,35] Ding *et al.* demonstrate a metasurface monolithically integrated on a silicon waveguide to create 2D beam focusing.[34] However, the use of gold-based meta-atoms is incompatible with the standard CMOS process. The resonance-based phase shifting mechanism also limits the precision of wavefront shaping because of the high sensitivity to the meta-atom dimensions.

In this paper, we report a monolithic and CMOS-compatible platform to integrate metasurfaces on silicon waveguides for versatile shaping of free-space emission. We demonstrate experimentally versatile wavefront shaping, including 2D diffraction-limited beam focusing with a Strehl ratio of 0.82 and holographic projection of an image above the chip. We fabricate the meta-atoms of amorphous silicon (a-Si) nanopillars on top of silicon waveguides to allow scalable monolithic integration. Simulation shows that the phase shift is introduced by the propagation through the nanopillar without relying on sensitive resonances, which enables reliable phase control in practice. We demonstrate that the meta-atoms evanescently coupled to a waveguide mode can convert the guided mode excitation to the free-space emission with well-determined phase shifts. The demonstrated metasurfaces can integrate with various waveguiding components to form a comprehensive chip-scale optical system for free-space applications.

We design metasurfaces monolithically integrated on standard 220 nm-thick silicon photonic waveguides for generating shaped emission, as shown in Figure 1a. The metasurfaces are composed of meta-atoms of 1.2 µm-tall a-Si nanopillars. The operating wavelength is 1550 nm. We arrange these nanopillars in a square lattice with a period Λ of 562 nm. These a-Si nanopillars couple to the $TE_0$ mode of the silicon photonic waveguide evanescently, perturbing the guided mode to create light emission to the free space. To ensure perturbative coupling strength, we introduce an 80 nm $SiO_2$ interlayer between the waveguide and the a-Si nanopillars. We perform 3D finite-difference time-domain (FDTD, Lumerical Inc.) simulation to study the intensity distribution in one of the evanescent-coupled meta-atoms in the lattice, as shown in Figure 1b. The intensity distribution is consistent with the eigenmode of a nanopillar in the lattice shown in Figure 1c. This indicates that an evanescent-coupled meta-atom in a lattice can be viewed as a small truncated cylindrical waveguide. Each nanopillar can introduce a phase shift to the emission, controlled by its effective refractive index $n_{\text{eff}}^{\text{pillar}}$ and thus by the nanopillar radius. This mechanism has been used in the literature to realize free-space polarization-independent metalenses.[39] We can thus engineer the emission from the waveguide and realize arbitrary phase profiles simply by varying the nanopillar radii at different positions. In contrast to conventional grating couplers in silicon photonics,[21,22] our metasurfaces allow phase assignment down to the unit cell level, enabling versatile wavefront shaping and holographic projection. We establish a library of meta-atom design parameters by performing simulation with the rigorous coupled-wave analysis (RCWA),[40] as shown in Figures 1d and 1e. With this library, we can look up the required geometrical design parameter at each position. RCWA allows a more efficient search of design parameters than the more rigorous but computation-intensive 3D FDTD. To simulate the evanescently-coupled nanopillar array in RCWA, we excite the nanopillars with a total internal reflected (TIR) plane wave source. The incident angle of the plane wave $\theta_i$ is chosen to be $\sin^{-1}(n_{\text{eff}}^{\text{slab}}/n_{\text{Si}})$ to produce the same evanescent field of the TE$_0$ slab waveguide mode, where $n_{\text{eff}}^{\text{slab}}$ is the effective refractive index (see Supporting Information). Figure 1d shows the phase of the emission from the nanopillar array as a function of the nanopillar height and radius. Considering the fabrication capabilities in our facility, we choose a height of 1.2 μm and radii ranging from 100 nm to 194 nm. As shown in Fig. 1e, this library provides a phase coverage of 1.9π, which is sufficient to produce almost any emission phase profile. We compare the emitting phase library obtained by RCWA with the more rigorous 3D FDTD simulation and obtain a good agreement. These results also match well with the phase response predicted by the eigenmode of the nanopillar (see Supporting Information). The agreement confirms the physical origin of the phase response: each evanescently-coupled nanopillar can be treated as a truncated cylindrical waveguide to produce a propagation phase delay determined by its radius.

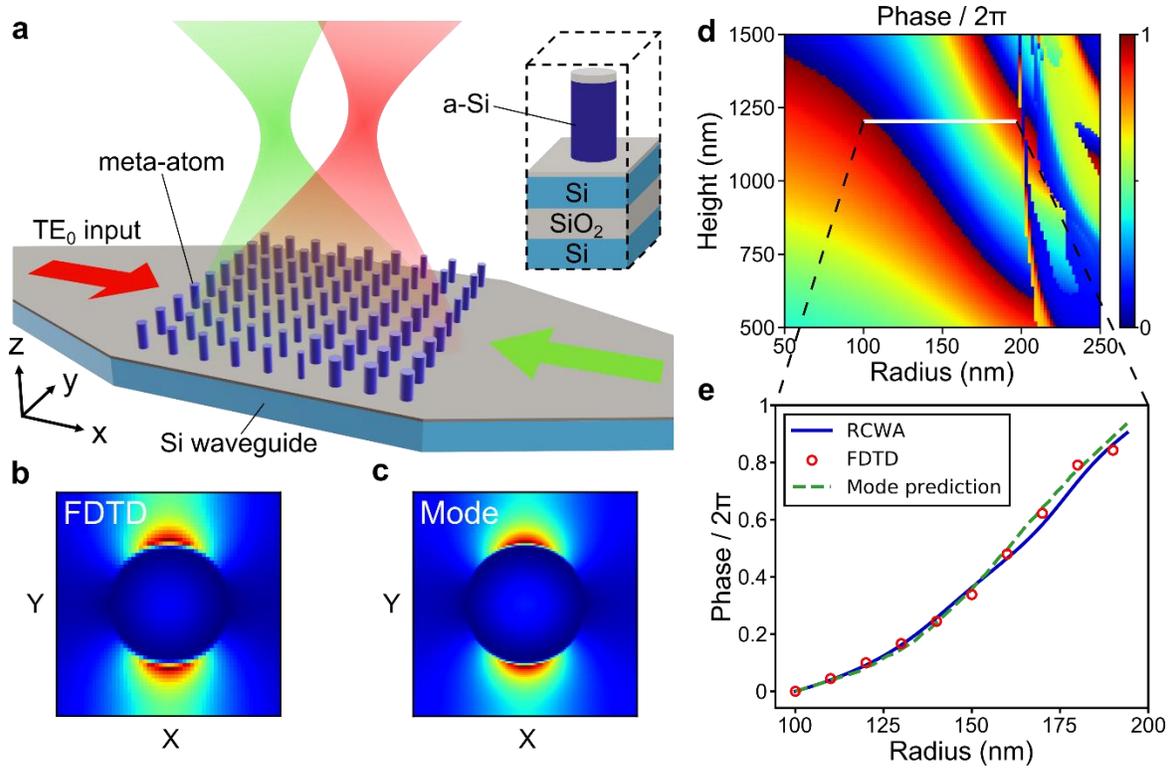

**Figure 1.** Schematic and simulation of the monolithic platform of metasurfaces on silicon photonics. (a) Schematic of a metasurface integrated on a silicon waveguide for switchable beam focusing. The metasurface consists of a-Si nanopillars in a square lattice. The metasurface is excited by the guided mode and creates focusing emission. The focused spot can be switched between two positions, depending on the excitation direction, as indicated by the green and red colors. Inset: schematic of the meta-atom unit cell (height: 1.2 μm; period: 562 nm). (b) FDTD-simulated intensity distribution near one of the nanopillars in the lattice when excited by the guided mode. The nanopillar radius is 150 nm. (c) Eigenmode of a nanopillar in the lattice (nanopillar radius: 150 nm). (d) RCWA-simulated phase of the emission from the nanopillar array as a function of the nanopillar height and radius. (e) Meta-atom phase library used in the design, where the radius ranges from 100 to 194 nm. We confirm the phase response with three different simulation methods: RCWA (blue solid line), 3D FDTD (red circle), and mode prediction (green

dashed line), showing the phase response originates from the propagation phase delay in each nanopillar.

With the phase library, we design a metasurface to produce focusing emission from a waveguide, where the focused spot position can be switched by controlling the excitation direction, as illustrated in Figure 1a. When the excitation guided mode propagates in the +x direction, it inherits the accumulating propagation phase $\beta x$, where $\beta$ is the propagation constant (Figure 2a). The phase profile $\phi_{em}(x, y)$ emitted to the free space can be expressed as $\phi_{em}(x, y) = \phi_{ms}(x, y) + \beta x$, where $\phi_{ms}(x, y)$ is the abrupt phase shift created by the metasurface. To produce a focused spot with a focal length $F$ normal to the metasurface, we choose the metasurface phase profile to be

$$\phi_{ms}(x, y) = -k_0\left(\sqrt{x^2 + y^2 + F^2} - F\right) - \beta x, \tag{1}$$

where the last term cancels with the propagating phase $\beta x$ to generate the ideal hyperbolic phase profile for the focusing emission.[24] Here we design a 60 μm × 20 μm metasurface with $F =100$ μm. This metasurface phase profile $\phi_{ms}(x, y)$ can be realized by looking up the proper meta-atom radii from the library. The corresponding metasurface pattern is shown by the scanning electron microscope (SEM) image of the fabricated metasurface in Figure 2c. The pattern is asymmetric in the x direction due to the last term in Eq. (1). The same metasurface produces a different emission phase profile when it is excited by the -x propagating guided mode, as depicted in Figure 2b. The emission phase profile becomes $\phi_{em}(x, y) = \phi_{ms}(x, y) - \beta x$. Using the grating equation $\beta = k_0 \sin\theta_d + 2\pi/\Lambda$, we can rewrite the emission phase profile as

$$\phi_{em}(x, y) = -k_0\left(\sqrt{x^2 + y^2 + F^2} - F\right) - 2k_0 \sin\theta_d\, x, \tag{2}$$

where $\theta_d$ is the emission angle for a grating. The last term creates a tilt to the wavefront. Therefore, given $\Lambda$=562 nm and $\beta= 2.8535k_0$ for the TE$_0$ mode, when we send the excitation guided mode in the -x direction, the focusing emission from the metasurface is tilted by -11°.

We perform 3D FDTD simulation to verify the switchable beam focusing from the metasurface and show diffraction-limited focusing quality. Figure 2d shows the emission intensity distribution on the xz plane when the metasurface is excited by the +x propagating guided mode. We observe a tightly focused spot at the height of 94 μm with a full width at half maximum (FWHM) spot size of 2.3 μm (x direction) by 8.2 μm (y direction). When we excite the metasurface with the -x propagating guided mode, as shown in Figure 2e, the emission is tilted by -11°, which is consistent with Eq. (2). The focused spot has an FWHM spot size of 2.4 μm (x direction) by 8.1 μm (y direction) at the height of 90 μm. The focusing efficiency for the +x propagating excitation case is 11 % (see Supporting Information for the definition), while 65 % of the input power passes through the metasurface-coupled region and remains in the waveguide. The focusing efficiency for the -x propagating excitation case is 14 %. In this case, 56 % of the input power remains in the waveguide after passing through the metasurface region. The focusing efficiency can be further improved by increasing the length of the metasurface in the x direction to extract all the waveguide power. Figure 2f and 2g show the cross-sections of the simulated intensity distributions at the focal plane along the x direction for the two excitation directions. We compare the simulated intensity distributions from the metasurface with the ideal intensity distributions produced by the uniform emission from a rectangular aperture of 60 μm × 20 μm with ideal hyperbolic phase profiles. We calculate the Strehl ratios from the comparison with the ideal intensity distributions (see Supporting Information). The simulated Strehl ratios are 0.85 and 0.83 for the +x and -x propagating excitation cases, respectively, indicating diffraction-limited performances. We analyze the mode purity of the light remaining in the waveguide after passing through the metasurface-coupled region. The intensity distribution is shown in Figure 2h. A fraction of 99 % remains in the $TE_0$ mode, indicating that the perturbation of waveguide-metasurface coupling does not excite any significant higher-order modes.

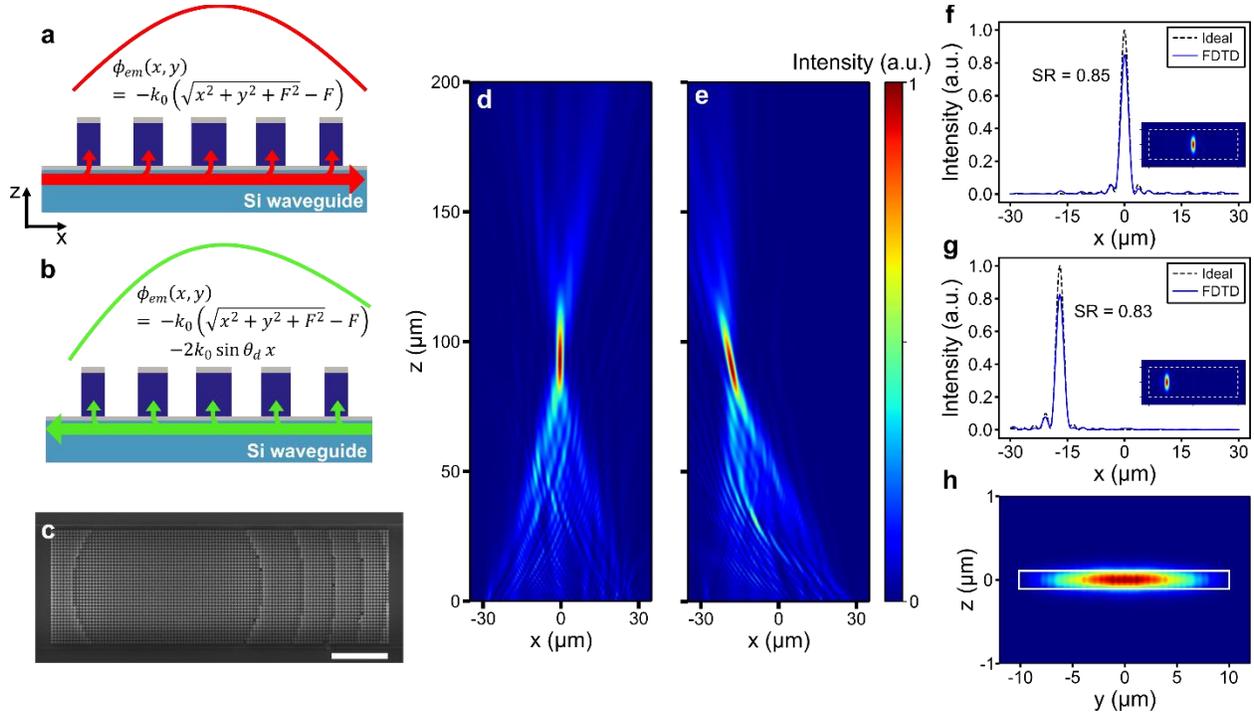

**Figure 2.** Design methodology and simulation of switchable beam focusing. (a,b) Illustration of the emission phase profile when the metasurface is excited by the +x (-x) propagating guided mode. (c) SEM image of the fabricated metasurface, showing the metasurface pattern for switchable beam focusing. Scale bar: 10 μm. (d,e) FDTD-simulated focusing intensity distribution on the xz plane when the metasurface is excited by the +x (panel d) and -x (panel e) propagating guided modes. The emission angle is tilted by -11° in panel e, showing a switchable operation. (f,g) Cross-sections of the FDTD-simulated intensity distributions at the focal plane when the metasurface is excited by the +x (panel f) and -x (panel g) propagating guided modes. For comparison, we also plot the ideal intensity distributions produced by the uniform emission from a rectangular aperture with ideal hyperbolic phase profiles. The simulated Strehl ratios are 0.85 and 0.83 in panels d and e, respectively. The intensity is normalized to the integrated power. Insets: 2D intensity distributions on the focal plane, where the white dashed line indicates an area of 60 μm × 20 μm. SR: Strehl ratio. (h) Simulated intensity distribution in the 22 μm × 220 nm silicon waveguide after the light passes through the metasurface-coupled region, in which 99 % of the power remains in the $TE_0$ mode.

We develop a process to fabricate the metasurface and the silicon waveguides using electron beam lithography (EBL) and a single plasma etching step, as shown in Figure 3a. First, we spin-coat electron-beam photoresist (ZEP 520A) on a silicon-on-insulator (SOI) substrate with a 3 μm buried oxide layer and a 220 nm top silicon layer. This is followed by the waveguide patterning using EBL. Electron beam evaporation is used to deposit 80 nm $SiO_2$, followed by the lift-off process to make the $SiO_2$ etch mask for the waveguides. After the waveguides are defined, we deposit 1.2 μm a-Si and 150 nm $SiO_2$ by plasma-enhanced chemical vapor deposition (PECVD). We spin-coat electron-beam photoresist (ma-N 2403) and pattern the metasurface structure with precise alignment. We use inductively-coupled plasma reactive-ion etching (ICP-RIE) to transfer the metasurface pattern to the $SiO_2$ etch mask using $C_4F_6$/Ar/$O_2$ based gases. After that, a single ICP-RIE etching step using HBr/$Cl_2$ based gases creates the metasurface and the waveguides together. Figure 3b shows the SEM image of the fabricated device.

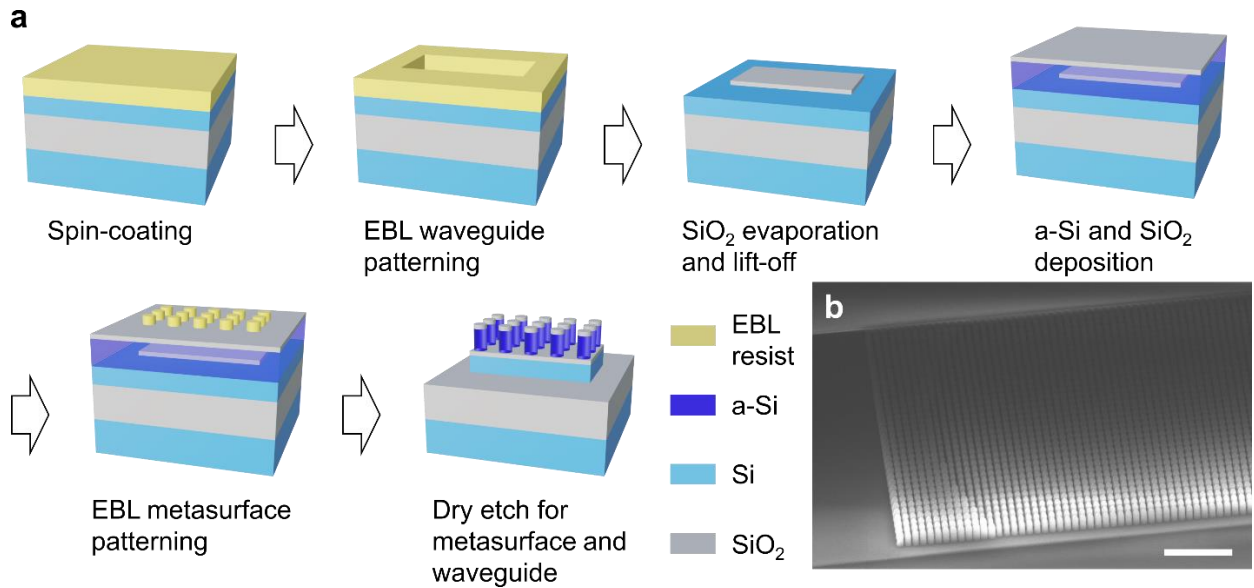

**Figure 3.** Fabrication processes and image of the devices. (a) Fabrication processes of the metasurface-on-waveguide structure using EBL and a single etching step. (b) Tilted SEM image of the metasurface monolithically integrated on a silicon waveguide. Scale bar: 5 μm.

We measure the emission from the monolithically-integrated metasurface and demonstrate switchable diffraction-limited beam focusing with a Strehl ratio of 0.82. The measurement setup is shown in Figure 4a. The light source is a tunable narrow-linewidth laser (Toptica CTL1550) operating at the wavelength of 1550 nm with a linewidth < 10 kHz. We couple the light from a cleaved single-mode fiber to the silicon photonic integrated circuit with a grating coupler. Light is routed by the silicon waveguides to the regions with the integrated metasurface, where light emits to the free space. The emission from the metasurface is captured by a home-built infinite-conjugate microscope with a 50X NA 0.65 objective (Mitutoyo M Plan Apo NIR) and an infrared camera (Xenics Bobcat 320). We translate the microscope with a motorized stage to image along different heights to map the 3D intensity distribution of the emission. Figures 4b and 4c show the measured intensity distributions on the xz plane when the metasurface is excited in the two different directions. When guided mode excitation is along the +x direction, a focused spot is formed vertically at z = 109 μm. As we switch to the opposite excitation direction, the focusing profile is titled by -10° and forms a spot at z = 102 μm. The slightly larger focal lengths and different tilt angle compared to the simulations are attributed to the meta-atom dimension errors in the fabrication. The cross-sections of the measured intensity distributions at the focal plane along the x direction are shown in Figures 4d and 4e, which agree with the FDTD simulations. The insets show the focused spot on the xy plane. In Figure 4d, the measured FWHM spot size is 2.1 μm (x direction) by 6.5 μm (y direction). The Strehl ratio is 0.82, indicating a diffraction-limited performance. In Figure 4e, the measured FWHM spot size is 2.2 μm (x direction) by 5.8 μm (y direction), and the Strehl ratio is 0.74. In the y direction, the measured FWHM spot sizes are smaller than the simulated values, which we attribute to the slightly distorted profiles along this direction (see Supporting Information).

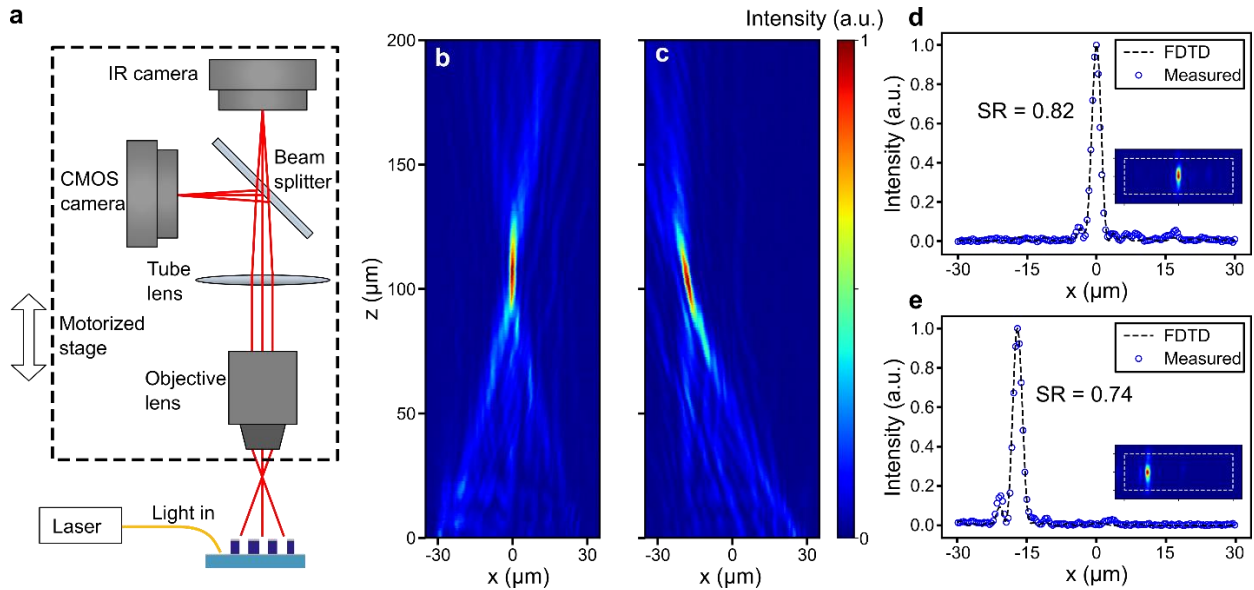

**Figure 4.** Measurement of the integrated metasurface for generating a switchable focusing emission profile. (a) Schematic of the measurement setup. (b,c) Measured intensity distributions on the xz plane when the metasurface is excited by the +x (panel b) and -x (panel c) propagating guided modes. The emission angle is tilted by -10° in panel c, showing a switchable operation. (d,e) Cross-sections of the measured and FDTD-simulated intensity distributions at the focal plane when the metasurface is excited by the +x (panel d) and -x (panel e) propagating guided modes. The peak intensity is normalized to 1. The measured Strehl ratios are 0.82 and 0.74 in panels d and e, respectively. Insets: 2D intensity distributions on the focal plane, where the white dashed line indicates an area of 60 μm × 20 μm. SR: Strehl ratio.

We realize an integrated meta-hologram experimentally to demonstrate the versatility of the platform of metasurfaces on silicon photonics. We design a meta-hologram with a size of 40 μm by 40 μm to project an image of the letter "ψ" at 20 μm above the metasurface. The required phase profile, as shown in Figure 5a, is found by the Gerchberg-Saxton algorithm.[41] We impart the required phase profile using the phase library in Figure 1e. We perform the full-wave 3D FDTD simulation to verify the design. Figure 5b shows that a simulated holographic image of the letter "ψ" is constructed at 20 μm above the meta-hologram. We use the same process shown in Figure 3a to fabricate the device. Figures 5c and 5d show the optical microscope (OM) and SEM images of the fabricated meta-hologram. The measured intensity distribution at 26 μm above the metasurface is shown in Figure 5e, which agrees well with the simulation. This demonstration shows the versatility of creating emission with a complicated wavefront beyond the capabilities of conventional grating couplers. The demonstrated on-chip holographic projection can find applications in near-eye display for virtual/augmented reality and single-pixel imaging.[11,42]

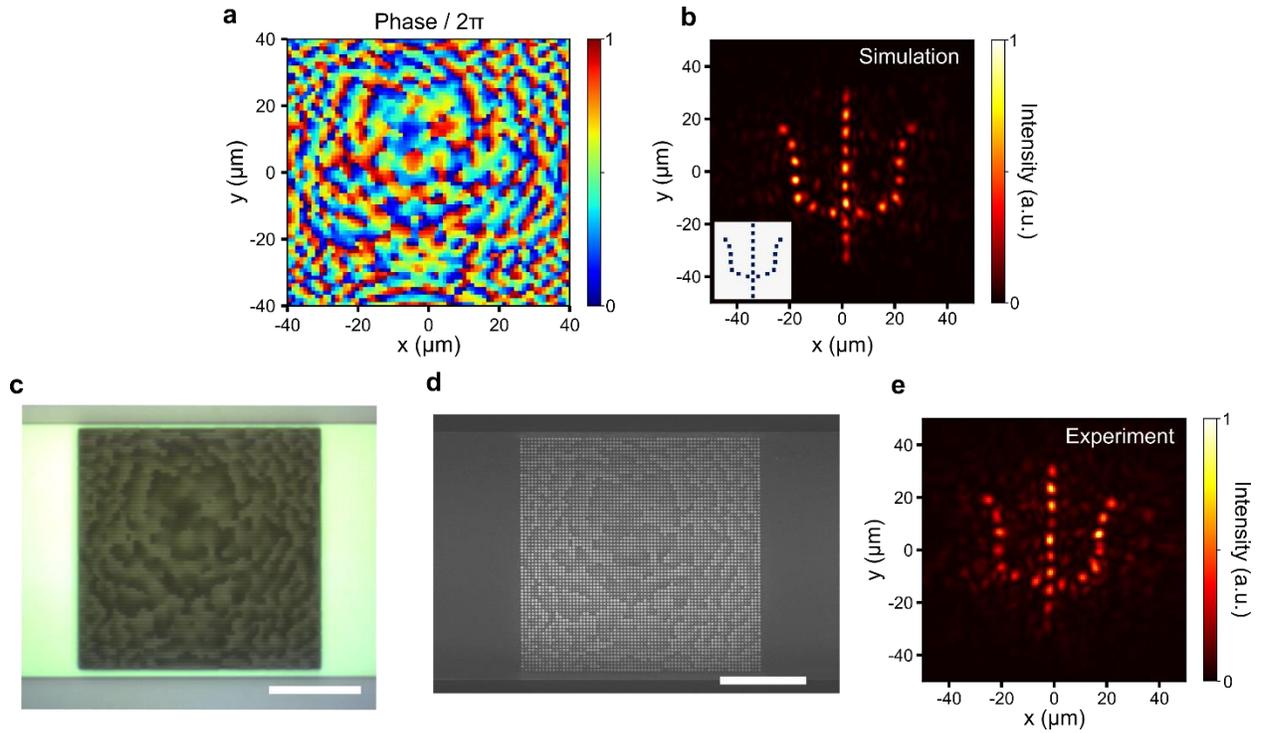

**Figure 5.** Simulation and experimental demonstration of an integrated meta-hologram. (a) Phase profile obtained by the Gerchberg-Saxton algorithm for projecting the letter "ψ" at 20 μm above the metasurface. (b) Simulated intensity distribution at 20 μm above the metasurface obtained by FDTD. Inset: target image. (c) OM image of the fabricated metasurface. (d) SEM image of the fabricated metasurface. Scale bar in panels c and d: 15 μm. (e) Measured intensity distribution at 26 μm above the metasurface.

In summary, we have demonstrated versatile shaping of free-space emission, including beam focusing and holographic projection, using a monolithic platform of metasurfaces on silicon photonics. The platform enables versatile wavefront control with a subwavelength resolution beyond the capabilities of conventional grating couplers. Given the meta-atom library, one can realize arbitrary emission phase profiles simply by the arrangement of meta-atoms without resorting to computationally intensive optimization[43,44]. This allows the realization of large-area integrated metasurfaces. Our demonstration at the telecommunication wavelength can potentially extend to the visible regime by constructing the metasurfaces with large-bandgap CMOS-compatible materials such as SiN and $TiO_2$.[45,46] The demonstrated integrated metasurfaces can be added to the existing silicon photonic ecosystem to provide a versatile interface between the free-space radiation and the guided modes. More free-space applications of silicon photonic integrated circuits can be envisioned with the offered extended functionalities.

## ASSOCIATED CONTENT

**Supporting Information**

The Supporting Information is available free of charge.

Detailed device structure; simulation of the phase library by RCWA; simulation of the phase and amplitude library by FDTD; modal analysis of a nanopillar in a lattice; focusing efficiency and Strehl ratio calculation; simulated and measured intensity cross-sections along the y axis (PDF)

## AUTHOR INFORMATION

**Corresponding Author**

You-Chia Chang

Department of Photonics, College of Electrical and Engineering, National Yang Ming Chiao Tung University, Hsinchu 30010, Taiwan; orcid.org/0000-0001-8842-2512;

Email: youchia@nycu.edu.tw

**Author Contributions**

P. Y. H., S. L. F., and Y. S. L. carried out the experiment. P. Y. H. performed the analysis. P. Y. H. and Y. C. C. wrote the manuscript. Y. C. C. conceived the original idea and supervised the project. P. Y. H. and S. L. F. fabricated the sample with support from W. H. H., J. M. S., and P. Yu.

**Funding Sources**

This work was supported by the Ministry of Science and Technology (MOST), Taiwan, R.O.C. under the award No. MOST 110-2222-E-A49-007- and MOST 110-2124-M-A49 -004-.

**Notes**

The authors declare no competing financial interest.

ACKNOWLEDGMENT

The authors thank Prof. Yao-Wei Huang in our department for fruitful discussions. This work was supported by the Ministry of Science and Technology (MOST), Taiwan, R.O.C. under the award No. MOST 110-2222-E-A49-007- and MOST 110-2124-M-A49 -004-. The authors would like to acknowledge fabrication support provided by Taiwan Semiconductor Research Institute (TSRI), Taiwan, R. O. C.

# Supporting Information for:

# Integrated Metasurfaces on Silicon Photonics for Emission Shaping and Holographic Projection

*Ping-Yen Hsieh,*[1] *Shun-Lin Fang,*[1] *Yu-Siang Lin,*[1] *Wen-Hsien Huang,*[2] *Jia-Min Shieh,*[1,2] *Peichen Yu,*[1] *and You-Chia Chang*[1,*]

[1]Department of Photonics, College of Electrical and Engineering, National Yang Ming Chiao Tung University, Hsinchu 30010, Taiwan

[2]Chip Integration Technology Department, Taiwan Semiconductor Research Institute, Hsinchu 30078, Taiwan

## 1. Detailed device structure of the metasurface on a silicon photonic waveguide

Figure S1 shows the schematic of a unit cell of the device. We fabricate the device on a silicon-on-insulator (SOI) wafer with a 220 nm top silicon layer and a 3 μm buried oxide layer. Between the metasurface and the silicon waveguide, we introduce an 80 nm $SiO_2$ layer for controlling the coupling strength. This layer also serves as the hardmask for the waveguide etching. The metasurface is composed of amorphous silicon (a-Si) nanopillars with a height of 1.2 μm. The nanopillars are arranged in a square lattice with a period of 562 nm. The radius of the pillars ranges from 100 to 194 nm. There is an 80 nm-thick SiO2 hardmask left on top of the a-Si nanopillar after metasurface etching.

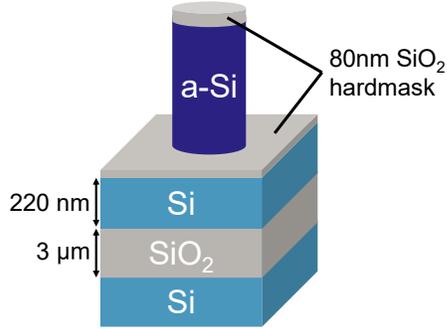

**Figure S1.** Schematic of a unit cell of the metasurface on the silicon waveguide.

## 2. Simulation of meta-atom phase library by rigorous coupled-mode analysis (RCWA)

We use RCWA to allow a more efficient search of the design parameters than the more rigorous but computation-intensive 3D finite-difference time domain (FDTD). Because RCWA only allows plane wave sources, we excite the nanopillars with a total internal reflected (TIR) plane wave source that produces the same evanescent field of the TE$_0$ slab waveguide mode (Fig. S2a). As shown by the simulation model in Fig. S2b, we send a plane wave source with $\lambda = 1550$ nm from a bulk silicon. The incident angle $\theta_i$ is determined by the condition $n_{Si} k_0 \sin \theta_i = n_{\text{eff}}^{\text{slab}} k_0$, where $n_{Si}$ is the refractive index of the bulk silicon, $k_0$ is the free-space wave number, and $n_{\text{eff}}^{\text{slab}}$ is the effective refractive index of the TE$_0$ mode of the waveguide with the nanopillars on top. In order to obtain the effective refractive index $n_{\text{eff}}^{\text{slab}}$ of a waveguide that includes a nanopillar array on its top, we perform a single FDTD simulation shown in Fig. S2c. We use the FDTD simulation to find the far-field diffraction angle $\theta_f$ and extract $n_{\text{eff}}^{\text{slab}}$ using the grating equation $k_0 \sin \theta_f = n_{\text{eff}}^{\text{slab}} k_0 - \frac{2\pi}{\Lambda}$, where $\Lambda = 562 nm$ is the period of the nanopillar array. The radius of the nanopillar is chosen to be a fixed value of 147 nm in determining $n_{\text{eff}}^{\text{slab}}$. We find the far-field diffraction angle $\theta_f$ to be 5.48°, which corresponds to $n_{\text{eff}}^{\text{slab}} = 2.8535$. Once we know $n_{\text{eff}}^{\text{slab}}$, we can calculate corresponding $\theta_i$ and use the RCWA simulation to calculate the emission phase library. The results are shown in Figures 1d and 1e in the main text.

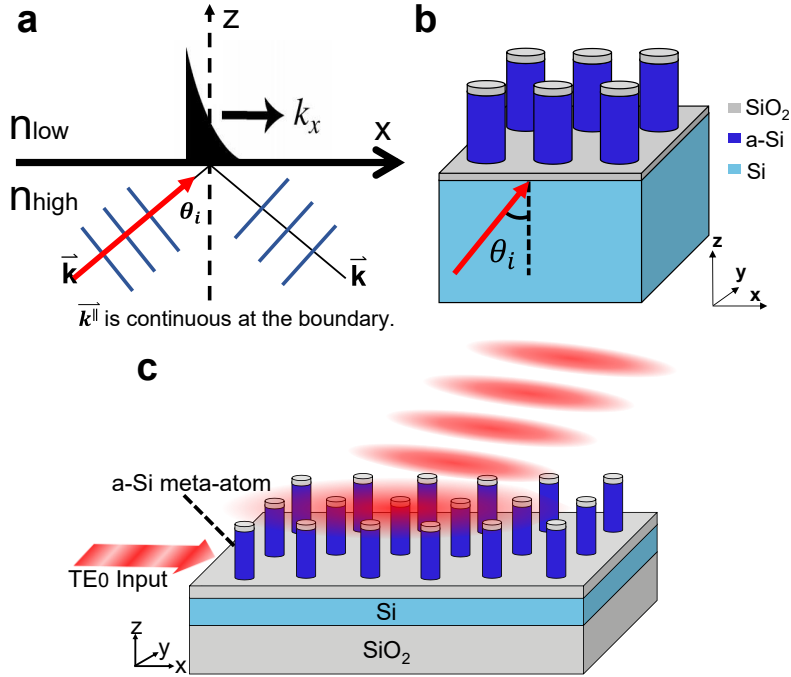

**Figure S2.** (a) Schematic of the generation an evanescent wave using a plane wave and the total internal reflection. (b) Schematic of the simulation model in RCWA. (c) Schematic of the simulation model in FDTD for determining the effective refractive index $n_{\text{eff}}^{\text{slab}}$. Periodic boundary condition is used in the y direction.

## 3. Simulation of meta-atom phase and amplitude libraries by finite-difference time-domain (FDTD)

We use the same FDTD simulation model described in the previous section (Figure S2c) to calculate the phase and amplitude libraries of the metasurface. We place a monitor above the meta-atom array and capture the complex emission electric field, including the amplitude and the phase. The procedure is repeated as we sweep through different radii. These results allow us to construct the phase and amplitude libraries.

A phase library provides a look-up table between the nanopillar radius and the phase shift, which is shown in Fig. 1e of the main text. To characterize the emission amplitude of the meta-atom, we use a quantity called the leakage parameter.[1] Leakage parameter characterizes the fraction of power that leaks from the waveguide per unit length, as defined by

$$-\frac{dP_{wg}}{dx} = 2[\alpha_{up} + \alpha_{down}]P_{wg}, \qquad (S1)$$

where $P_{wg}$ is the power remaining in the waveguide. $\alpha_{up}$ and $\alpha_{down}$ are the leakage parameters for the upward and downward emission, respectively. In the FDTD simulation, we extract the leakage parameters by monitoring the power leaking to the free space and the power remaining in the waveguide. Figure S3 shows the leakage parameter $\alpha_{up}$ as a function of the nanopillar radius, which represents the amplitude library of the metasurface.

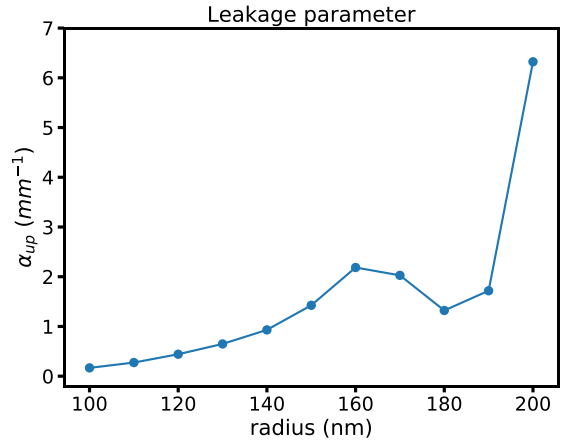

**Figure S3.** Amplitude library of the metasurface obtained by the FDTD simulation, which shows the leakage parameter $\alpha_{up}$ as a function of the nanopillar radius.

## 4. Modal analysis of the nanopillar

Each Si nanopillar can be viewed as a truncated cylindrical waveguide.[2] Here we predict the phase shift associated with the eigenmode when the light passes through the Si nanopillar. We use the eigenmode solver (MODE solutions, Lumerical Inc.) to solve the fundamental eigenmode of a silicon nanopillar array arranged in a square lattice with a period = 562nm. We calculate the effective refractive index $n_{eff}^{pillar}(r)$ for different nanopillar radius $r$. The phase shift after passing through the nanopillar $\phi(r) = k_0 n_{eff}^{pillar}(r) h$, where $k_0$ is the free-space wave number, and $h$ is the height of the nanopillar. The result is shown in Figure 1e in the main text.

## 5. Focusing efficiency calculation

The focusing efficiency is defined as

$$\text{Focusing efficiency} = \frac{\text{Power that reaches the focused spot}}{\text{Input power that enters the waveguide}}. \tag{S2}$$

We calculate the power that reaches the focused spot by integrating the intensity over a rectangular area on the focal plane. The widths of the rectangular area are chosen to be 6 times the full-width at half-maximum (FWHM) spot sizes in the x and y directions.

## 6. Strehl ratio calculation

We quantify the focusing quality by the Strehl ratio. We calculate the ideal intensity distribution by simulating with FDTD the uniform emission from a rectangular aperture of 60 μm

× 20 μm with the ideal hyperbolic phase profile given by $\phi(x,y) = -k_0(\sqrt{x^2 + y^2 + F^2} - F)$. We obtain the ideal intensity distribution on the focal plane $I_{\text{ideal}}(x,y)$, which follows well the sinc² function. To calculate the Strehl ratio of the measured data, we capture the intensity distribution on the focal plane $I_{\text{measure}}(x,y)$. Both $I_{\text{ideal}}(x,y)$ and $I_{\text{measure}}(x,y)$ are normalized to their integrated power, where the integration is taken over a rectangular area with widths of 6 times the FWHM spot sizes in the x and y directions. The ratio between the peak intensities of $I_{\text{measure}}(x,y)$ and $I_{\text{ideal}}(x,y)$ gives the Strehl ratio. A Strehl ratio higher than 0.8 is conventionally considered diffraction-limited.

## 7. Simulated and measured intensity cross-sections along the y axis

In the main text, we show the simulated and measured intensity cross-sections along the x axis in Figures 2f, 2g, 4d, and 4e. Here we show the results along the y axis. Figures S4a and S4b show the FDTD-simulated intensity cross-sections at the focal plane along the y direction for the two excitation directions. The FDTD-simulated FWHM spot sizes are 8.2 μm and 8.1 μm in Fig. S4a and S4b, respectively. For comparison, we also plot the ideal intensity cross-sections produced by the uniform emission from a rectangular aperture of 60 μm × 20 μm with ideal hyperbolic phase profiles. Figures S4c and S4d show the measured intensity cross-sections at the focal plane along the y direction. The measured FWHM spot sizes are 6.5 μm and 5.8 μm in Fig. S4c and S4d, respectively.

We notice that the measured FWHM spot sizes along the y direction are smaller than the FDTD-simulated sizes, as shown in Fig. S4c and S4d. However, the measured profiles show shoulder-like distortions. These results show that the FWHM spot size does not fully characterize the performance of the focusing performance. Strehl ratio is a better indicator because it is associated with the integrated power over an area. However, the information of the entire intensity distribution $I(x,y)$ is needed to show all the details of the focusing performance.

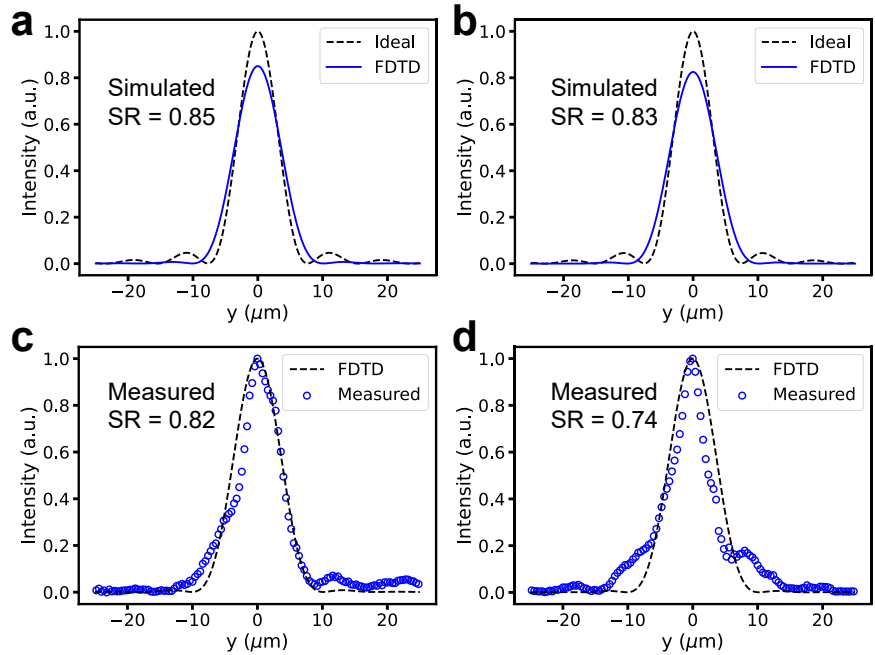

**Figure S4.** The simulated and measured intensity cross-sections along the y direction. (a,b) Cross-sections of the FDTD-simulated intensity distributions at the focal plane along the y direction when the metasurface is excited by the +x (panel a) and -x (panel b) propagating guided modes. For comparison, we also plot the ideal intensity distributions produced by the uniform emission from a rectangular aperture with ideal hyperbolic phase profiles. The intensity is normalized to the integrated power. (c,d) Cross-sections of the measured and FDTD-simulated intensity distributions at the focal plane along the y direction when the metasurface is excited by the +x (panel c) and -x (panel d) propagating guided modes. The data is normalized to the peak intensity. SR: Strehl ratio.